\documentclass[twocolumn]{aastex63}
\usepackage[utf8]{inputenc}
\usepackage{comment}
\usepackage[title]{appendix}
\usepackage{amsmath}
\usepackage{booktabs}
\usepackage{enumerate}
\usepackage[para]{threeparttable}

\begin{document}
\title{Templates for Fitting Photometry of Ultra-High-Redshift Galaxies} 

\author[0000-0003-3780-6801]{Charles L. Steinhardt}
\affiliation{Cosmic Dawn Center (DAWN)}
\affiliation{Niels Bohr Institute, University of Copenhagen, Lyngbyvej 2, K\o benhavn \O~2100, Denmark}

\author[0000-0002-5588-9156]{Vasily Kokorev}
\affiliation{Cosmic Dawn Center (DAWN)}
\affiliation{Niels Bohr Institute, University of Copenhagen, Lyngbyvej 2, K\o benhavn \O~2100, Denmark}

\author[0000-0001-7633-3985]{Vadim Rusakov}
\affiliation{Niels Bohr Institute, University of Copenhagen, Lyngbyvej 2, K\o benhavn \O~2100, Denmark}
\affiliation{Cosmic Dawn Center (DAWN)}

\author[0000-0002-1740-9641]{Ethan Garcia}
\affiliation{Cosmic Dawn Center (DAWN)}
\affiliation{California Institute of Technology, 1200 E. California Blvd., Pasadena, CA 91125, USA}

\author[0000-0002-5460-6126]{Albert Sneppen}
\affiliation{Niels Bohr Institute, University of Copenhagen, Lyngbyvej 2, K\o benhavn \O~2100, Denmark}
\affiliation{Cosmic Dawn Center (DAWN)}

\begin{abstract}

Recent data from the James Webb Space Telescope allow a first glimpse of galaxies at $z \gtrsim 11$.  The most successful tool for identifying ultra-high-redshift candidates and inferring their properties is photometric template fitting.  However, current methods rely on templates derived from much lower-redshift conditions, including stellar populations older than the age of the Universe at $z > 12$, a stellar initial mass function which is physically disallowed at $z > 6$, and weaker emission lines than currently observed at $z > 7.5$. Here, two sets of synthetic templates, optimized for the expected astrophysics of galaxies at $8 < z < 12$ and $z > 12$, are developed and used to fit three galaxies at $z > 12$ from the SMACS0723 field.  Using these improved templates, quantitative estimates are produced of the bias in inferred properties from JWST observations at $z>8$ due to these effects. The best-fit redshifts are similar to those found with previous template sets, but the inferred stellar masses drop by as much as 1--1.6 dex, so that stellar masses are no longer seemingly inconsistent with $\Lambda$CDM.  The two new template sets are released in formats compatible with \textsc{EAZY} and \textsc{LePhare}. 
\end{abstract}


\section{Introduction}
\label{sec:intro}

The launch of the James Webb Space Telescope (JWST) provides the first observational window into galaxies at $z \gtrsim 11$.  Within a few weeks of the first data release, several galaxies have been found at $z > 12$ \citep{Adams2022,Atek2022,Castellano2022,Donnan2022, Finkelstein2022, Naidu2022,Yan2022}.  The discovery of these galaxies and an analysis of their properties will allow tests of models for halo assembly and early galaxy evolution, and will likely be one of the key early results from JWST.

Photometric surveys have been the best discovery tool for high-redshift galaxies \citep{Steidel1992}, and all of the early, ultra-high-redshift sources have been detected in NIRCam imaging.  Both the identification of these galaxies and the subsequent determination of properties such as stellar mass ($M_*$) and star formation rates (SFR) rely on photometric template fitting.  One of several codes \citep{Brammer2008,LePhare,Leja2017,Carnall2018} is used to compare a series of model spectral templates against the observed photometry, and the redshift and properties are inferred from the best-fit models.  Most models are synthetic templates produced from stellar population synthesis under a range of assumptions and conditions. Crucially, all these synthetic templates are derived from local conditions, which are either theoretically disallowed or observationally disproven for galaxies at ultra-high redshift.   

Therefore, in this work, a new template set is developed with the aim of producing models tuned for fitting ultra-high-redshift galaxies.  These templates are also available for download at \url{https://github.com/e-m-garcia/hot-templates} in formats compatible with \textsc{EAZY} and \textsc{LePhare}.  The models are significantly different in three respects. Firstly, they use younger stellar populations since at those redshifts, the stellar population cannot be more than a few hundred Myr old. In contrast, all current templates are older than the age of the universe at $z\approx12$ \citep{Brammer2008}. Secondly, we apply bottom-lighter IMFs corresponding to expected conditions in star-forming regions of galaxies at $z \sim 15$ (see \S~\ref{sec:imf}). Thirdly, we use stronger emission lines as none of the local strong-line metallicity calibrations provide a good prediction of the observed metallicities at $z>7.5$ \citep{Curti2022}. The differences in inferred parameters between these templates and ones derived from current templates are discussed in \S~\ref{sec:params}.  

Throughout this paper we assume a flat $\Lambda$CDM cosmology with $\Omega_{\mathrm{m},0}=0.3$, $\Omega_{\mathrm{\Lambda},0}=0.7$ and H$_0=70$ km s$^{-1}$ Mpc$^{-1}$.

\section{The High-Redshift Initial Mass Function}
\label{sec:imf}

The physical parameters inferred for high-redshift galaxies will depend upon the assumptions made when producing the synthetic spectra used to fit them.  Under the correct assumptions, templates should produce both a better fit and a more accurate description of the key physical properties of the galaxy than with incorrect assumptions.  One central assumption common to all current templates is a universal stellar initial mass function (IMF), which is assumed to be identical to the current Galactic IMF.  This is a critical assumption for inferring properties, because the light emitted by a galaxy is dominated by the most massive stars comprising only a small fraction of the stellar mass, and the IMF is necessary to infer the remainder of the stellar population from that high-mass tail. However, theoretical models for star formation predict that the IMF should depend upon several properties of star-forming molecular clouds, including gas temperature and metallicity \citep{LyndenBell1976,Larson1985,Jermyn2018,Steinhardt2020}.  Thus, assuming a universal IMF is, in effect, an assumption that the differences between Galactic and extragalactic star-forming regions are negligible.

The ideal approach would be to construct templates using a range of possible initial mass functions and determine which one fits best.  This is the approach taken in recent work fitting galaxies at lower redshifts \citep{Sneppen2022,Steinhardt2022a,Steinhardt2022b}.  Based on theoretical arguments \citep{Jermyn2018}, a one-parameter family of templates at different gas temperatures $T_g$ was produced.  The corresponding IMF at each gas temperature was taken as 

\begin{equation}
    \frac{dN}{dm} \propto 
    \begin{cases}
       m^{-0.3} &\quad m < 0.08 M_{\odot} \cdot \bigg( \frac{T_g}{T_0} \bigg)^2 \\
       m^{-1.3} &\quad 0.08 M_{\odot} \cdot \bigg( \frac{T_g}{T_0} \bigg)^2 < m < 0.5 M_{\odot} \cdot \bigg( \frac{T_g}{T_0} \bigg)^2 \\
       m^{-2.3} &\quad 0.5 M_{\odot} \cdot \bigg( \frac{T_g}{T_0} \bigg)^2 < m, \\
    \end{cases}
    \label{eq:IMF}
\end{equation}
which matches a Galactic (Kroupa) IMF at $T_g = T_0 = 20$ K.  Other investigations of the relation between temperature and IMF produce a similar relationship, but with a slightly different temperature dependence \citep{Jeans1902,Hopkins2012,Chabrier2014}.  Each of these sets of templates was used to fit photometric observations from the COSMOS2015 catalog \citep{Laigle2016}, and the best-fit among all of the template sets was taken as having the correct IMF and used to infer physical parameters.

However, as there are strong spectral covariance between changes in the IMF and changes in dust or metallicity, it was only possible to constrain the IMF for the $\sim 10$\% of the catalog with the highest signal-to-noise ratio in \cite{Sneppen2022}.  Further, the COSMOS catalog includes multi-wavelength measurements in up to 30 bands.  Thus, a similar approach used on NIRCam photometry, with typically 4-5 bands exhibiting non-zero flux emitted by the most distant observable galaxies, it is not possible to observationally constrain the IMF.  Rather, it will be necessary to simply choose an IMF (or, equivalently, extinction and metallicity) and assume it holds for all galaxies in the sample.  Indeed, this is already being done by all template fitting codes, which assume that the high-redshift IMF is the same as the Galactic one. 

An alternative might be to produce observational rather than synthetic templates.  Using NIRSpec, it should be possible to take spectra of a number of these $z \sim 15$ sources, and those with confirmed high redshifts can then be used as templates to fit others.  A similar empirical approach is taken for one of the template sets in \textsc{LePhare} \citep{LePhare}.  However, the spectrum alone will not be able to constrain the IMF, and thus this approach cannot be used to infer properties such as stellar mass or star formation rate.  If the sole goal is redshift determination, this may still be the best source of templates, as the templates will certainly correspond to at least some types of $z \sim 15$ galaxies.

The remaining approach, then, is to simply assume a $z \sim 15$ IMF and hope that the assumption is approximately correct.  This is already being done with the assumption of a Galactic IMF, and the goal here is to improve upon that assumption. There are three key observational and theoretical indications validating a change of IMF in this ultra-high-redshift regime:  

First, the best-fit IMFs for COSMOS galaxies exhibit several clear trends \citep{Sneppen2022}.  Quiescent galaxies at all redshifts out to $z \sim 2$ are well fit with approximately-Galactic IMFs \citep{Steinhardt2022b}.  However, for star-forming galaxies, there is a characteristic, best-fit IMF at each redshift, which becomes bottom-lighter towards high redshift \citep{Steinhardt2022a}.  Inferred values of $T_g$ rise from 25 K at low redshift to over 30 K by $z = 4$.  An extrapolation of this characteristic $T_g$ to $z \sim 15$ produces an estimate of $35-40$ K. 

Second, a reasonable interpretation is that in star-forming galaxies, the young stellar population is the dominant contribution to gas temperature in star-forming regions.  If so, $z \sim 15$ galaxies, which must have very high SFR in order to have formed so quickly, should have even higher $T_g$ and even bottom-lighter IMFs.  A closer examination of the first stages of star formation, taking place primarily in galactic centers, finds $T_g$ of $45-60$ K (Steinhardt et al., in prep.), with 60 K characteristic of galaxies in their earliest stages of evolution.  

Finally, at high redshift the CMB temperature is an additional source of heating.  At $z \sim 15$, the CMB temperature is 44 K, and if those stars have been forming over the past $\sim 120$Myr, a typical age for galaxies on the star-forming main sequence, a luminosity-weighted CMB temperature at the time of star formation would be closer to 55 K.  It is difficult to find mechanisms for star-forming clouds to cool well below the CMB temperature, so this should be a minimum possible value of $T_g$ at high redshift.

All three of these estimates suggest that a reasonable value of $T_g$ for galaxies at $z \sim 10$ would be $40-50$ K and at $z \sim 15$ significantly higher.  In this work, the choice is made to develop templates based on $T_g = 60$ K for use when modeling the highest-redshift galaxies in JWST at $z > 12$.  This produces a significantly bottom-lighter IMF than in the Milky Way.  An additional set of templates with $T_g = 45$ K is also  provided, and is likely a better model for galaxies at $8 < z < 12$.

In summary, inferring properties from JWST observations creates somewhat of a quandary.  Ultra-high redshift galaxies with very young stellar population are simultaneously (1) the galaxies for which inferred properties are most sensitive to the choice of IMF; (2) the galaxies most likely to form stars under very different conditions than in our own Galaxy today; and (3) the galaxies for which it will be most difficult, if not impossible, to constrain the IMF observationally.  

Fortunately, measurements from large surveys at lower redshift do provide constraints on the IMF, and their measured IMFs are consistent with physical intuition.  Thus, the choice made in this work is to use similar physical intuition to extrapolate to the most likely ultra-high-redshift IMF, and this is almost certainly a better approximation to the true IMF than a Galactic one.  However, it would not be shocking to discover that this physical intuition breaks down under the extreme conditions in the earliest galaxies, and inferred properties such as stellar masses and star formation rates for ultra-high-redshift galaxies should be assumed to have systematic uncertainties well in excess of those produced using the same techniques at lower redshift.

\section{Fits and Inferred Parameters}
\label{sec:params}

The IMFs described in this work are implemented for two different photometric template fitting codes, \textsc{EAZY} \citep{Brammer2008} and \textsc{LePhare} \citep{LePhare}.  \textsc{EAZY} starts from a set of 560 model spectra generated using Flexible Stellar Population Synthesis (FSPS), which span a wide range of SFR, stellar population ages, metallicities, and extinctions, then condenses them into 12 basis vectors which approximately span the full space\footnote{For some catalogs, additional templates are added to produce an expanded basis.}.  Within that limited basis, matrix inversion allows a rapid determination of the best-fit linear combination of those 12 templates to the observed photometry over a grid of redshifts.  \textsc{LePhare}, instead, relies on a grid search.  For a smaller number of star formation histories, a grid of ages, redshifts, and extinctions is compared against the observed photometry to produce the best fit.  \textsc{LePhare} will therefore only consider physically reasonable models, although the grid search means that often only a coarse search is possible in a reasonable runtime and the best-fit model might not be found.  \textsc{EAZY} can find the best fit more quickly within the limited basis, but in principle the best-fit linear combination might be physically unreasonable if the true parameters lie well outside of that basis.  

Implementation for a grid-based search such as \textsc{LePhare} is straightforward, as it merely requires generating templates for every combination under consideration at a variety of ages.  Instructions for installing the template set in this work for use with \textsc{EAZY} (or other codes that use the same format, e.g., \textsc{Stardust}; \citealt{kokorev2021}) and
\textsc{LePhare} can be found in Appendices \ref{app:eazy} and \ref{app:lephare}, respectively.  The remainder of this section focuses on the template sets produced for \textsc{EAZY}, which additionally require a reconsideration of the template basis.

\subsection{Modifying the Standard Template Basis}

The standard \textsc{EAZY} release includes a set of 12 basis templates with properties chosen to approximately span the full observed galaxy population in large photometric catalogs.  A list of the templates and their properties can be found in Table \ref{table:proper_orig}.  The standard \textsc{EAZY} templates are based on a Chabrier IMF, but the IMFs used here (Eq. \ref{eq:IMF}) are instead equivalent to a Kroupa IMF at $T_g = 20$ K.  Therefore, a set of templates using a standard Kroupa IMF is used for comparison in order to demonstrate the effects of varying the IMF on inferred properties.

\begin{table}
    \label{tab:templates}
    {\centering
    \caption{Properties of the stellar populations in the 12-template basis of \textsc{EAZY}.  The properties are nearly identical for different IMFs, as they are constructed from linear combinations of FSPS models with the same physical parameters.}
    \label{table:proper_orig}
    \begin{tabular}{c|c|c|c}
        \hline\hline
        ID & Age (Gyr) & $A_{V}$ & $Z$  \\ \hline
        0 & 0.72 & 0.06 & 0.0066 \\ 
        1 & 3.15 & 0.11 & 0.0119 \\ 
        2 & 10.59 & 0.13 & 0.0140 \\ 
        3 & 13.83 & 0.12 & 0.0223 \\ 
        4 & 13.64 & 0.73 & 0.0237 \\ 
        5 & 15.12 & 2.88 & 0.0240 \\ 
        6 & 0.34 & 0.07 & 0.0154 \\ 
        7 & 0.62 & 0.27 & 0.0211 \\ 
        8 & 1.07 & 0.80 & 0.0216 \\ 
        9 & 1.30 & 1.37 & 0.0216  \\ 
        10 & 2.26 & 1.81 & 0.0217 \\ 
        11 & 2.44 & 2.94 & 0.0216 \\ \hline
    \end{tabular}}
\end{table}
Each template is a specific linear combination of a broader set of 560 individual models.  To produce templates at $T_g = 45$ K, each of the 560 individual models was run with a modified IMF as in Eq. (\ref{eq:IMF}).  Those templates were then combined into the 12 basis templates using the same coefficients.  This produces a very similar set of luminosity-weighted average properties for each template (see Table \ref{tab:properties}).  Because of the bottom-lighter IMF, the light in each template is dominated by slightly younger stars which have produced a slightly higher extinction and metallicity than at $T_g = 20$ K.

\begin{figure}
    \centering
    \includegraphics[width=0.49\textwidth]{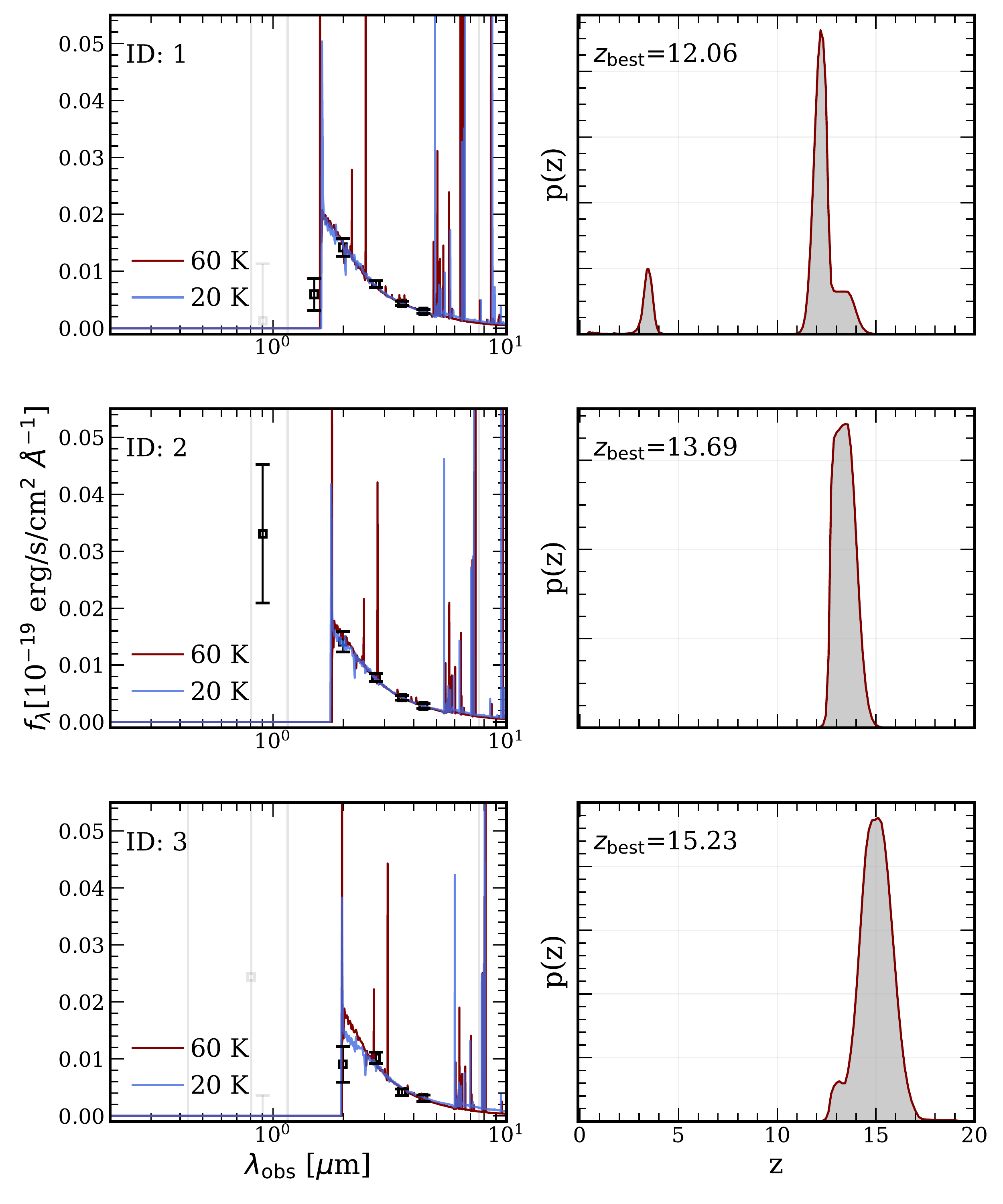}
    \caption{Best-fit reconstructed spectra (left) and redshift probability functions (right) for three galaxies in SMACS0723. The fits were performed with the standard \textsc{EAZY} Chabrier IMF at $T_g = 20$K (blue) and with Kroupa IMF at $T_g = 60$K (maroon), with the latter fit expected to be a better description of galaxies at $z > 12$.}
    \label{fig:sed_fits}
\end{figure}

For illustration, these templates have been used to fit three high-z candidates ($z>12$) from the SMACS0723 cluster field (Kokorev et al. in prep.). These have been labelled as SMACS 1, 2 and 3.  The data reduction of the raw images and the photometry extraction have been performed with the \textsc{grizli} pipeline \citep{grizli}. The images have been additionally aligned to match the astrometric reference frame used for the ALMA Lensing Cluster Survey data reduction of \textit{HST} data \citep{Kokorev2022}. The data reduction and the photometric catalog will be fully described in Brammer et al. 2022 (in prep.). The best-fit redshifts using the standard \textsc{EAZY} templates are $z = 12.23$, 13.48 and 15.09, respectively.  The 60 K Kroupa IMF templates produce a very similar redshift, which is not surprising given that these are dropout galaxies in the $JWST$ F090W band  and the redshift is primarily determined from the presumed location of the break.  The fits themselves are also broadly similar, with typical differences of $\lesssim 2$\% in the reconstructed flux (Fig. \ref{fig:sed_fits}).

\begin{figure}[b]
    \centering
    \includegraphics[width=\linewidth]{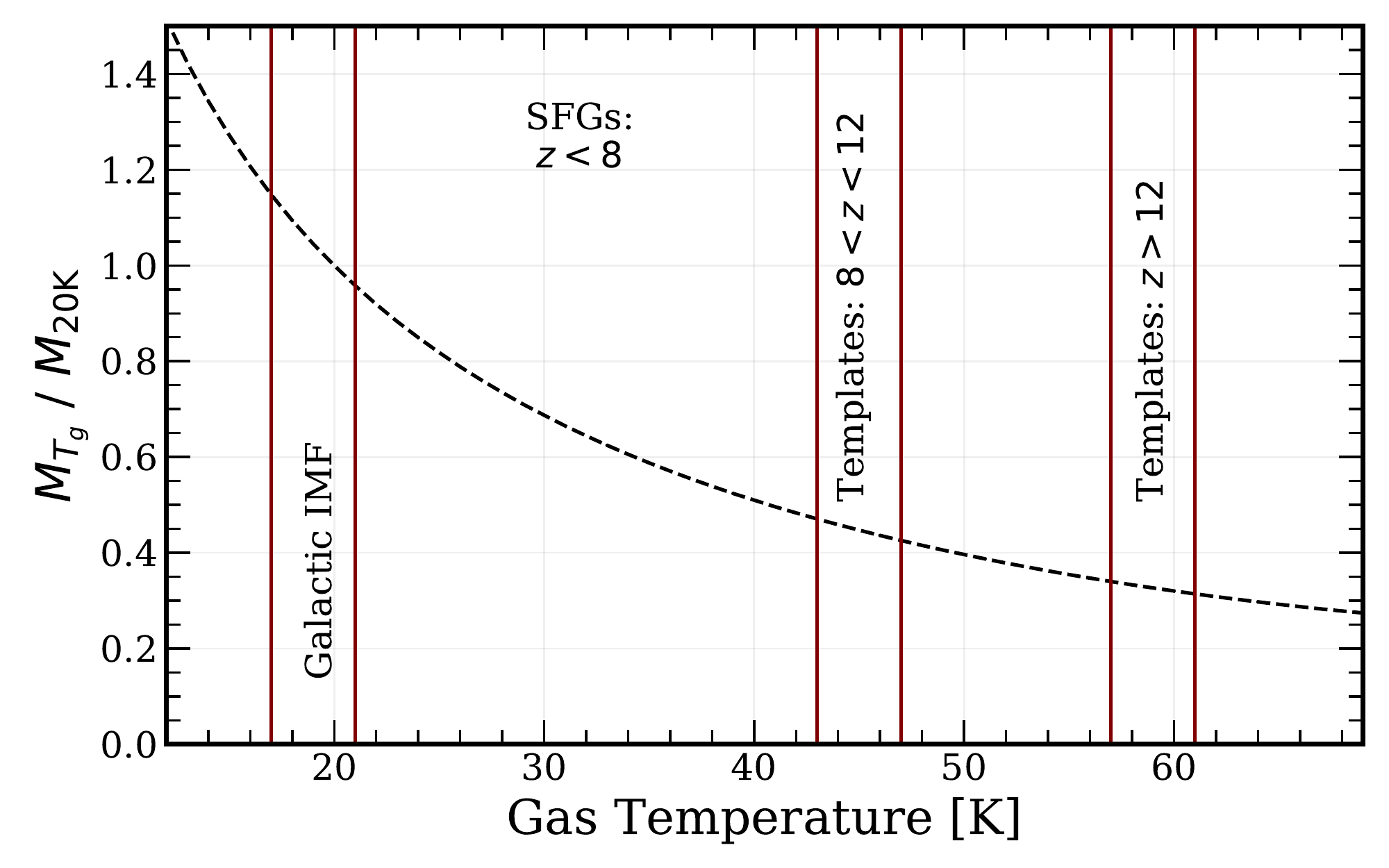}
    \caption{Expected change in inferred stellar masses using the IMFs in this work relative to Galactic IMF as a function of $T_g$. The dashed black line indicates the average shift based on the analytical integral while holding all other parameters fixed, which is a close match for the observed differences in fitting the three $z > 12$ candidates under the same conditions.  The actual shift in stellar mass will combine this effect, the difference between Kroupa and Chabrier choices of Galactic IMF, and changes in stellar age and nebular emission lines and will typically exceed the analytical integral shown here.}
    \label{fig:IMF_bias}
\end{figure}

The physical properties exhibit a far greater difference (see Fig. \ref{fig:IMF_bias} or Table \ref{tab:properties}).  In a star-forming galaxy, most of the light comes from the most massive stars, while most of the stellar mass lies in the least massive ones.  Thus, a significantly bottom-lighter IMF will produce a much lower stellar mass and SFR.  

\begin{table*}
\begin{center}
\begin{threeparttable}[t]
\caption{Physical properties for the best-fit template reconstruction of three galaxies in the SMACS0723 field.  Four template sets are compared: (1) the standard \textsc{EAZY} basis, derived using a Chabrier IMF; (2) the same basis derived with a Kroupa IMF; (3) the same basis at $T_g = 60$K; and (4) a new $T_g = 60$ basis with younger stellar populations and other properties tuned for ultra-high-redshift galaxies.  All template bases agree that these sources lie at $z > 12$, but the inferred masses differ far more significantly.  There is insufficient information to use a goodness-of-fit test to determine which of these template sets will fit best, so the approach recommended here instead relies on astrophysical assumptions about these galaxies.}
\label{tab:properties}
\centering
\begin{tabular}{ccccccc}
\hline \hline
IMF & Template Set & $\log_{10}(M/ M_{\odot})$ & SFR ($M_\odot~\rm{yr}^{-1}$) & $t$ (Gyr) & $A_V$ & $z$ \\ 
\hline
\multicolumn{6}{c}{SMACS 1 (07:22:44.85, -73:29:53.75)}\\
\hline \hline
Chabrier (20 K) & \textsc{EAZY}  12 & $8.85^{+0.00}_{-0.00}$ \tnote{a} & 3.02  & 0.34 & 0.07 & 12.23  \\ 
Kroupa (20 K) & \textsc{EAZY} 12 & $8.64^{+0.08}_{-0.20}$ & 4.28  & 0.38 & 0.07 & 12.41  \\ 
Kroupa (60 K) & \textsc{EAZY}  12 & $8.13^{+0.12}_{-0.24}$ & 2.45  & 0.39 & 0.07 & 12.41 \\ 
Kroupa (60 K) & Modified 6 & $8.04^{+0.12}_{-0.14}$ & 2.13  & 0.20 & 0.10 & 12.06 \\ \hline
\hline\multicolumn{6}{c}{SMACS 2 (07:22:51.93, -73:29:21.89)} \\\hline \hline
Chabrier (20 K) & \textsc{EAZY}  12 & $8.91^{+0.00}_{-0.00}$ & 3.47  & 0.34 & 0.07 & 13.48 \\ 
Kroupa (20 K) & \textsc{EAZY}  12 & $8.60^{+0.02}_{-0.02}$ & 3.32  & 0.34 & 0.07 & 14.26 \\ 
Kroupa (60 K) & \textsc{EAZY}  12 & $8.06^{+0.02}_{-0.02}$ & 1.54  & 0.33 & 0.07 & 14.30 \\ 
Kroupa (60 K) & Modified 6  & $8.18^{+0.15}_{-0.12}$ & 1.78  & 0.13 & 0.13 & 13.69 \\ \hline
\hline\multicolumn{6}{c}{SMACS 3 (07:23:03.45, -73:28:47.01)}\\\hline \hline
Chabrier (20 K) & \textsc{EAZY}  12 & $9.00^{+0.00}_{-0.00}$ & 4.29  & 0.34 & 0.07 & 15.09 \\
Kroupa (20 K) & \textsc{EAZY}  12 & $8.70^{+0.03}_{-0.03}$ & 4.25  & 0.34 & 0.07 & 15.77 \\
Kroupa (60 K) & \textsc{EAZY}  12 & $8.16^{+0.03}_{-0.03}$ & 1.97  & 0.33 & 0.07 & 15.80 \\ 
Kroupa (60 K) & Modified 6  & $7.79^{+0.16}_{-0.12}$ & 1.72 & 0.10 & 0.01 & 15.23\\ \hline
\end{tabular}
\begin{tablenotes}
\item[a] Uncertainties shown here are those reported from the \textsc{EAZY} covariance estimation.  As described in \S~\ref{subsec:errors}, these fit uncertainties significantly underestimate the true uncertainties for reconstructions consisting only of one template.  Thus, negligible uncertainties do not indicate a very well-constrained fit, but rather that the true SED lies well outside of the template basis.
\end{tablenotes}
\end{threeparttable}
\end{center}
\end{table*}
The best-fit stellar masses are between 0.5 and 0.6 dex lower at $T_g = 60$ K than at $T_g = 20$ K.  At lower redshifts, differences in the best-fit extinction and metallicity create a more complex relationship between $T_g$ and the best-fit stellar mass \citep{Sneppen2022,Steinhardt2022b}.  However, galaxies at $z \sim 15$ have not had enough time to form as much dust.  Thus, this change in stellar mass is very close to the difference in mass-to-light ratios for a single stellar population between $T_g = 20$ K and 60 K (Fig. \ref{fig:IMF_bias}).  A similar shift exists in the best-fit star formation rates.

\subsection{Younger Stellar Templates for High-Redshift Galaxies}
\label{subsec:errors}

With either set of fits, the inferred specific star formation rates (sSFR, or SFR/$M_*$) are $\sim 10^{-8}$/yr\footnote{In the standard \textsc{EAZY} basis, a higher sSFR cannot be inferred, since none of the 12 basis templates has an sSFR over $\sim 10^{-8}$/yr.}.  At that rate, the entire stellar population would form in 100 Myr.  At $z = 12.4$, 14.3, and 15.7, the Universe is 350 Myr, 290 Myr, and 260 Myr old, respectively.  Thus, 100 Myr is a plausible age for a stellar population in a $z \sim 15$ galaxy.   

However, every one of the basis templates corresponds to an older stellar population than in these high-redshift galaxies.  The template with the youngest stellar population age, template 6 (see Table \ref{table:proper_orig}), has a luminosity-weighted age of 340 Myr.  Nevertheless, as the youngest stellar population, it is the closest match.  The best-fit linear combinations of the 12 \textsc{EAZY} templates are dominated by this single youngest template, with template 6 contributing at least $\sim 76 \%$ to the fits of the object 1 and $100 \%$ to the other fits.  The next-youngest template, template 7, contributes all of the remainder.  The true SED very likely lies outside of the basis spanned by these 12 templates.

If the true SED lies well outside of the basis spanned by these 12 templates, the closest match will likely lie at an extreme edge of that basis.  This is consistent with the \textsc{EAZY} best-fit reconstructions, which consist almost solely of a single template.  If measurement uncertainties perturb the input photometry, it will still result in a galaxy younger than any of the basis templates, and so the best fit will still be dominated by the same single template.  Thus, even a large uncertainty in photometry will result in a minimal change in the best-fit spectrum and parameters.  
  
Because the uncertainties on \textsc{EAZY} fit parameters are derived from the uncertainty on the best fit within this 12-template basis, the estimated uncertainty on parameters such as stellar mass will also be unreasonably low.  The \textsc{EAZY} uncertainty estimation in the stellar masses for the three SMACS galaxies using the standard template basis breaks, as the covariance matrix cannot be calculated.  The resulting reported zero uncertainties are not an indication of reliability in the masses, but rather of how far outside of the template basis the observed photometry apparently lies.

Here, a set of six additional templates is introduced to solve this problem.  They include three different stellar population ages and two choices of extinction at each age, all of the same metallicity (Table \ref{table:proper}).  Two of these templates also have increased equivalent widths of emission lines to $\rm{EW(H\beta} + $[O {\sc iii}]$\lambda 5007) \approx 1370$~\AA.  Such extreme nebular emission, thought to be associated with the ``IRAC excess'' in galaxies at $z>7$, was observed in IRAC photometry (eg., \citealp{smit2014,oesch2015,roberts_borsani2016}), NIRCam photometry \citep{Labbe2022}, as well as NIRSpec spectra \citep{Carnall2022}.
The six extended templates $T_g = 45$ K and $60$ K are available online, with installation instructions in Appendix \ref{app:eazy}.

\begin{table}
    \label{a}
    \centering
    \caption{Properties of the stellar populations in the 6 new FSPS templates for \textsc{EAZY} at $T_g = 45$ K and 60 K.  The templates all have young stellar populations with ages between 0.05 and 0.20 Gyr for 45 K and between 0.1 and 0.5 Gyr for 60 K.  The low extinction and metallicity are expected to be typical of ultra-high-redshift galaxies.  Because of the strong emission lines present in early NIRSpec observations  \citep{Carnall2022}, two templates are included with strongly enhanced EW(H$\beta$ + [O {\sc iii}]$\lambda5007$) with respect to the standard FSPS setting of $\sim 274~\AA$.}
    \begin{tabular}{c|c|c|c|c|c}
    \hline\hline
        ID & $t_{45}~(\rm{Gyr})$ & $t_{60}~(\rm{Gyr})$ & $A_{V,45,60}$ & $Z_{45,60}$ & Neb. Em. \\ \hline
        1 & 0.1 & 0.05 & 0.005 & 0.0012 & 1x (std.) \\
        2 & 0.1 & 0.05 & 0.5 & 0.0012 & 5x \\
        3 & 0.3 & 0.1 & 0.005 & 0.0012 & 1x (std.) \\
        4 & 0.3 & 0.1 & 0.5 & 0.0012 & 1x (std.) \\
        5 & 0.5 & 0.2 & 0.005 & 0.0012 & 1x (std.) \\
        6 & 0.5 & 0.2 & 0.5 & 0.0012 & 5x \\ \hline
    \end{tabular}
    \label{table:proper}
\end{table}

The two different values for extinction of the new templates are $A_{V} = 0.005$ and $A_{V} = 0.5$.  Template 6, the dominant template in the fits with the standard \textsc{EAZY} basis with all three IMFs considered, has $A_V = 0.06$, which is therefore the extinction fit for all three galaxies using those bases.  Although the fits with the new six-template basis could produce extinction as high as $A_V = 0.5$, all three SMACS objects are fit with much lower extinctions similar to Template 6.  

Although large multi-wavelength surveys can have as many as 30 photometric bands, which can be parameterized by a basis of 12 templates, these ultra-high-redshift candidates will typically will only have detections in 4-5 bands. Thus, the basis we advocate using for JWST objects is necessarily smaller than the full EAZY basis.

Finally, the ability of the new templates to accurately constrain redshift was cross-checked by fitting an object in the relevant redshift range detected spectroscopically with NIRSpec, which was first reported in \cite{Carnall2022} (object ID 4590 in the paper).  The best-fit redshift at $T_g=45$ K is $z_{phot}=8.32$, which agrees with $z_{spec}=8.498$ well within one $\sigma$ of the $z_{phot}$ distribution.
  
\section{Discussion}
\label{sec:discussion}

This paper presents a revised and improved basis of templates for fitting high-redshift galaxies with $z>8$. While current synthetic templates are derived from stellar populations, which are impossibly old, have weaker emission lines than observations and invoke an unphysical assumption of a Galactic IMF, the revised templates improve on all these accounts.  Here, a new set of six templates optimized for the extreme conditions in the earliest galaxies is presented, along with instructions for use with \textsc{EAZY} and \textsc{LePhare}. 

It should be noted that the addition of these templates to the standard 12 \textsc{EAZY} basis templates will provide additional degrees of freedom, and JWST observations alone will typically not provide enough information to constrain the best fit within a combined, 18-template basis.  Further, the inclusion of very young stellar populations might allow spurious fits for some lower-redshift galaxies, since in combination with dusty or quiescent templates, it allows matching the blue and red ends of the SED nearly independently.  For similar reasons, deriving a best-fit IMF for each individual source is not possible, so it is instead necessary to select a single set of templates based on the expected astrophysics of typical very high-redshift galaxies.

Therefore, to avoid improving high-$z$ fits and properties at the cost of degrading low-$z$ objects we recommend a two-stage procedure.  First, the standard \textsc{EAZY} template basis should be used to fit all sources, since it is well optimized for the local objects in a photometric survey, which constituted the vast majority.  Then, the objects with best-fit $z > 8$ should be refit with these new templates: the template set at $T_g = 45$ K at $8 < z < 12$ and with $T_g = 60$K at $z > 12$.  The latter fits should produce better estimates of the physical parameters for these galaxies, which are robustly fitted at the highest redshifts. The new templates introduced are intended as a first-order improvement to current low-redshift derived templates, but are certainly not a complete solution. Future spectroscopy at z$>$12 may provide improved properties for template-fitting at these redshifts.

\subsection{Effects on Redshift Determination}

For galaxies at $z < 4$ in COSMOS, a change in IMF had negligible effect on the best-fit redshifts \citep{Sneppen2022}.  Most objects have detected flux on both sides of the Balmer break, often from narrow bands, and therefore the location of the break was well constrained.  However, the highest-redshift JWST galaxies are instead found as dropouts in NIRCam photometry, with no narrow bands to further constrain the location of the break.  The bluest bands will have no detection, the reddest bands will have clear detections, and one band in between (F200W at $z \sim 15$) will typically have a fainter detection, combining part of the spectrum above the Lyman break with negligible flux at shorter wavelengths.  The strongest constraint on the best-fit redshift comes from essentially calculating the equivalent width for that partial detection and determining what fraction of this band has non-zero flux.  Thus, a change in the UV slope near the Lyman break will change that equivalent width and unlike at $z < 4$, the corresponding best-fit redshift will be sensitive to the IMF.

For example, with everything else held constant,the switch from a Chabrier IMF to a Kroupa IMF within the standard \textsc{EAZY} template basis yields a higher best-fit redshift.  Because the best-fit models with this 12-template basis are solely composed of the same single template with both IMFs, a higher redshift is found for each SMACS source.  These changes in redshift are still within the estimated redshift probability density function, which describes the statistical uncertainty.  However, they are not statistical, but rather a systematic offset between the redshifts calculated from each template set due to smaller equivalent widths with a Kroupa IMF.    %

With the six-template basis proposed in this work, a more complex dependence is possible.  In addition to a change in IMF, the additional templates allow galaxies to be fit with different metallicities and extinction values than using the standard \textsc{EAZY} basis.  Thus, in total the UV slope of the reconstructed spectrum could be either steeper or shallower, and thus the redshift could change in either direction.  

\subsection{Impossibly Early Galaxies}

The strongest effect will be on the inferred stellar masses.  The change between a Chabrier and Kroupa IMF can produce a difference of $0.2-0.3$ dex in stellar mass, but the effect of a high $T_g$ is far stronger with variations of $\sim 0.5-1$ dex.  

This change is particularly relevant at very high redshift, because the inferred stellar masses from standard template libraries are high enough to challenge the standard $\Lambda$CDM paradigm.  Prior to JWST, the most massive, highest-redshift galaxies in large photometric surveys were already difficult to reconcile with $\Lambda$CDM halo mass functions \citep{Steinhardt2016,Behroozi2018}, an effect described as the impossibly early galaxy problem \citep{Steinhardt2016}.  One possible solution would be an increase in the stellar baryon fraction \citep{Finkelstein2015,Behroozi2020b}, which would allow halos of the same mass to contain galaxies with greater stellar masses at high redshift.  However, the estimated masses of JWST galaxies at $z = 10$ \citep{Labbe2022} are high enough that even this is no longer a solution: if the inferred stellar masses and redshifts are correct, the stellar masses would exceed the full baryon masses within halos at $z = 10$ \citep{BoylanKolchin2022}.  

The $T_g = 45$ K templates should provide better fits for $z = 10$ galaxies than templates with a Galactic IMF.  However, these templates cannot satisfactorily match the observed photometry, since these $z = 10$ candidates were selected because they appear to exhibit strong Balmer breaks.  None of the templates in the 6-template basis have sufficiently strong Balmer breaks, which take $\gtrsim 300$ Myr to develop for a single stellar population and $\gtrsim$ 500 Myr for any of the more realistic star formation histories used in these templates.  The age of the Universe at $z = 10$ is less than 500 Myr.  \citet{Labbe2022} solve this problem by using an older and dustier stellar population than the $T_g = 45$ K template set to produce a strong Balmer break.  This reconstructed stellar population is significantly older than 500 Myr and thus physically impossible.  Given current models, it is not possible to produce as strong of a Balmer break at $z = 10$ as exists in the reported photometry.

Perhaps the most likely explanation is that improved zero point corrections will significantly reduce the strength of the break.  Another possibility is that these are very dusty, much lower-redshift sources.  Constraining the redshifts to assume these are truly $z \sim 10$ sources and using the $T_g = 45$ K templates, the stellar masses for typical $z = 10$ galaxies drop by 1.6 dex.  This large difference comes from a combination of several effects:   (1) a bottom-ligher IMF; (2) the difference between Kroupa and Chabrier IMFs; (3) a younger stellar population; (4) lower extinction; and (5) enhanced nebular emission lines producing a fainter continuum.

This last effect could provide an alternative explanation for the apparent Balmer breaks in these $z = 10$ galaxies.  NIRSpec observations find stronger emission lines at $z > 7.5$ \citep{Carnall2022,Harikane2022} than are present in FSPS templates \citep{fsps,Conroy2010} derived from lower-redshift observations.  Sufficiently strong emission lines might even mimic a spectral break.  If so, followup spectroscopy of these massive $z \sim 10$ candidates might instead find very strong emission lines but no discernible Balmer break.  This would be consistent both with theoretical expectations for a stellar population when the Universe is less than 500 Myr old and with the observed combination of strong emission lines without Balmer breaks in the \citet{Carnall2022} NIRSpec spectroscopy.

\begin{figure}[b]
    \centering
    \includegraphics[width=\linewidth]{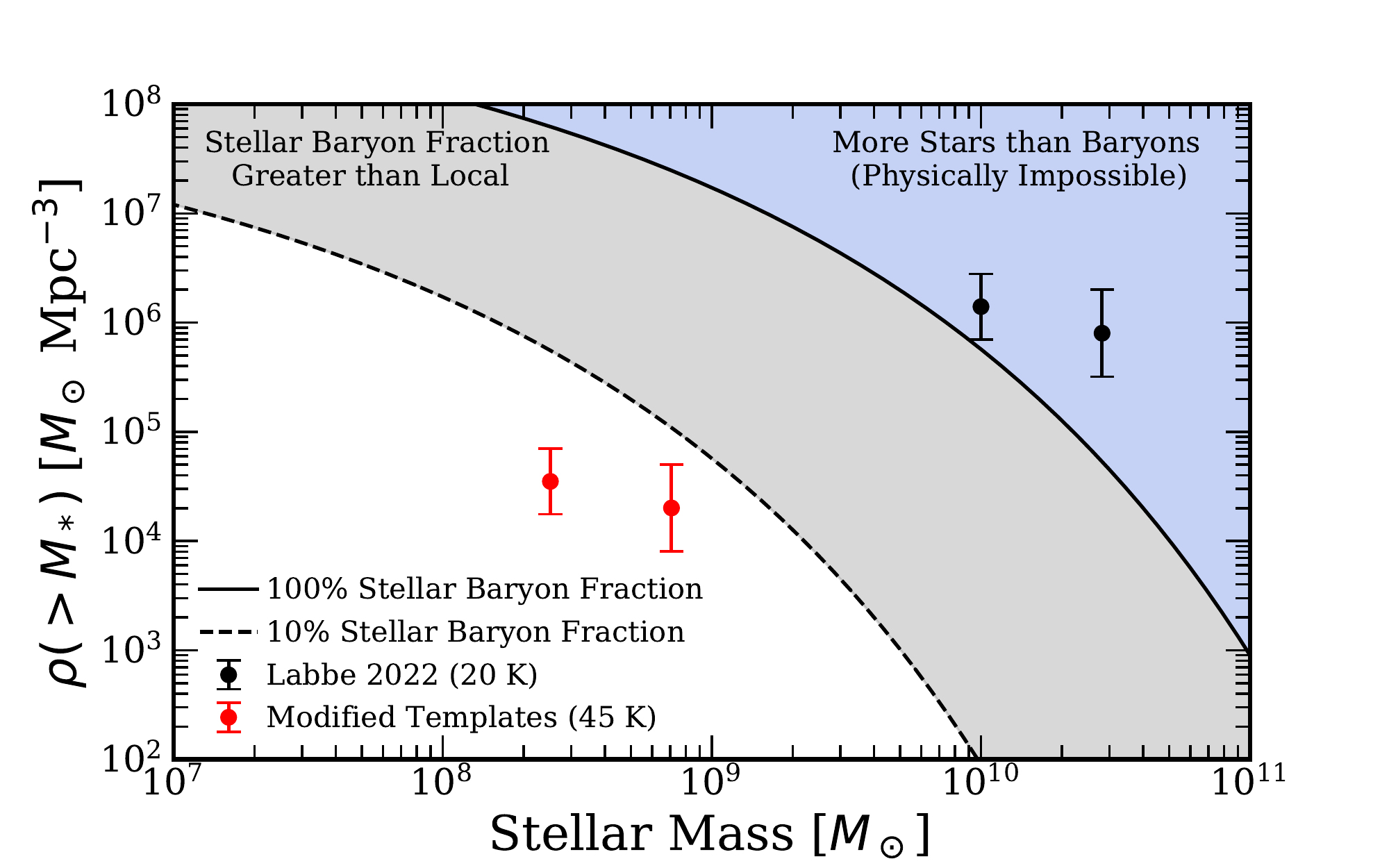}
    \caption{A switch from the standard template basis to the ones developed in this work reduces the best-fit stellar masses.  Using stellar masses derived by \citet{Labbe2022} from modified $T_g = 20$ K \textsc{EAZY} templates (black), \citet{BoylanKolchin2022} find that $z = 10$ galaxies are too massive to have been produced with a $\Lambda$CDM cosmology (this figure is reproduced from Fig. 2 in \citet{BoylanKolchin2022}).  However, the lower masses from templates optimized for high-redshift galaxies (red) can still be consistent with halo formation under the standard cosmological paradigm even with a similar stellar baryon fraction to the local Universe.}
    \label{fig:hmfvssmf} 
\end{figure}

Although the Galactic IMF-derived stellar masses are inconsistent with $\Lambda$CDM, the $T_g = 45$ K masses can still be reconciled with theoretical mass functions (Fig. \ref{fig:hmfvssmf}).  More generally, the high-luminosity tail of the luminosity function at very high redshift would no longer necessarily correspond to stellar masses as high as at more moderate redshifts.  As selection improves, it is possible that additional massive objects will be discovered, and the stellar mass densities estimated at present might be underestimated.  If so, a complete survey might still find tension between theoretical models for halo formation and observed high-redshift galaxies.  However, at present the impossibly early galaxy problem can be solved solely with a change to a template basis consistent with CMB heating.

Spectroscopy and improved modeling will ultimately be needed to determine whether these galaxies are truly at the stellar masses inferred from template fitting with any of these bases.  Currently, however, one must assume an IMF in order to determine physical parameters, which are essential in order to use JWST observations to test fundamental astrophysics and cosmology.  Given the extreme conditions at very high redshifts, it is necessary to use templates derived from a physically-motivated IMF rather than one derived from local observations under conditions that are physically impossible at high redshift.  

The authors would like to thank Michael Boylan-Kolchin, Gabriel Brammer, Andreas Faisst, Johan Fynbo, Kate Gould, Bahram Mobasher, Sune Toft, and Darach Watson for helpful comments.  The Cosmic Dawn Center (DAWN) is funded by the Danish National Research Foundation under grant No. 140.

\bibliographystyle{mnras}
\bibliography{refs.bib} 

\begin{appendices}
\section{Installing Templates in \textsc{EAZY}}
\label{app:eazy}

\textsc{EAZY} templates constructed for high redshift studies consist of two two sets of 6 templates.  Each set has the IMF corresponding to gas temperatures $T_g=45$ and $60$ K.  They were produced using FSPS \citep{fsps,Conroy2010}, with the 6 new templates representing stellar populations younger, more metal poor than in the standard 12 basis \textsc{EAZY} set and with stronger nebular emission (see Table~\ref{table:proper}).

For more detailed information on usage and further resources, refer to the Github repository at \url{https://github.com/e-m-garcia/hot-templates}.  \textsc{EAZY} with the standard set of templates is available at \url{https://github.com/gbrammer/eazy-py}.

\section{Installing Templates in \textsc{LePhare}}
\label{app:lephare}

The set of templates produced for usage consist of 4 templates at 45 K and 60 K each of varying star-formation histories at a z=0.0012 metallicity. These are reconstructed in FSPS \citep{fsps,Conroy2010} which allows for replacing the original Chabrier IMF \citep{Chabrier2003} with a temperature-dependent Kroupa IMF \citep{Kroupa2001, Jermyn2018}.  The templates are written into a BC03 ASCII file format such that \textsc{LePhare} can read them.

To use the templates, simply include the given templates in the galaxy template directory in \textsc{LePhare}, and use the young stellar age list included. It is also recommended to start from parameters associated with the existing 12 BC03 templates used for COSMOS when reading the templates, constructing the magnitude grid, and estimating redshifts and properties, although these can be modified to fit the needs of the fitting (such as different filters). For more detailed information on usage and further resources, refer to the Github at \url{https://github.com/e-m-garcia/hot-templates} and the \textsc{LePhare} repository at \url{https://gitlab.lam.fr/Galaxies/LEPHARE}.

\end{appendices}

\label{lastpage}
\end{document}